# Sub-seasonal Modulation and Predictability of Indian monsoon hourly Rainfall Extremes

Bijit Kumar Banerjee[1,2], Devabrat Sharma[3,4], Mahen Konwar[5], Simanta Das[6], Utpal Sarma[1,2] and B. N. Goswami[2*]

[1]Department of instrumentation and USIC, Gauhati University, Guwahati-781014, Assam, India
[2]ST radar Centre, Gauhati University, Guwahati-781014, Assam, India
[3]Department of Aerospace Engineering, Indian Institute of Technology Madras, Chennai-600036, India
[4]Centre of Excellence for Studying Critical Transitions in Complex Systems, Indian Institute of Technology Madras, Chennai-600036, India
[5]Indian Institute of Tropical Meteorology, Pune-411008, India
[6]Department of Physics, Cotton University, ,Guwahati-781001,Assam,India

*Corresponding author Email: bhupengoswami100@gmail.com

## 1 Abstract

Hourly rainfall extremes cause some of the most destructive weather disasters, yet numerical weather prediction models still struggle to forecast them, and a physical basis for their predictability remains unclear. Here, we identify a trivariate clustering of hourly rainfall extremes with surface temperature, phases of the Monsoon Intraseasonal Oscillation (MISO), and precipitable water vapor, establishing a physical foundation for the medium range predictability of these events. This clustering arises from multiscale interactions in which extremes organize into storm systems embedded within mesoscale convective clusters and synoptic low-pressure systems during active MISO phases. We develop an algorithm to identify, track, and monitor these storm systems. Although rapid error growth limits the prediction of isolated hourly extremes, our results provide basis for a physics informed training of deep learning, data driven models to forecast organized clusters of hourly rainfall extremes more than a week in advance, offering substantial potential to reduce losses from extreme rainfall.

## 2 Introduction

Approximately 80% of annual rainfall over India occurs during the Indian summer monsoon between June and September[1]. Climatological mean rainfall is highly inhomogeneous, with Northeast India and the western flank of the Western Ghats receiving heavy precipitation, while northwestern India remains semi-arid[2]. During the monsoon season, the most destructive rainfall events in tropical regions often occur on hourly time scales, producing cloudbursts, landslides, flash floods, and urban inundation. Both the frequency and intensity of these events are expected to rise rapidly in a warming climate[3–5]. Hourly extremes originate from sub-mesoscale convective systems with lifetimes of 3 to 6 hours. Yet most studies of rainfall extremes over India rely on daily gridded datasets[6,7], which represent mesoscale systems with spatial scales of 100 to 200 km and temporal scales of about one day. As greenhouse gas emissions continue to warm the planet, reliable estimates of the frequency and intensity of hourly extreme rainfall events (HEREs) are essential for disaster preparedness, loss reduction, and climate adaptation. In monsoon-dominated countries such as India, the societal risk posed by HEREs is especially high. Recent assessments suggest that socio-economic losses from weather extremes in India could reach nearly 10% of GDP under a 3°C rise in global mean temperature[8,9].

The India Meteorological Department maintains a large rain gauge network and has produced two gridded daily rainfall datasets spanning 1901 to the present[6,7]. In contrast, stations with hourly rainfall records are few[10], and those with long, continuous records are even fewer (Fig. 1H). The short duration of available records and the large interannual variability of hourly extremes make trend estimation unreliable. In the absence of robust hourly observations, most previous studies have focused on daily extremes and their spatial variability using datasets of varying lengths[7,11–15]. Given the devastating societal impacts of HEREs, reliable forecasts of their intensity and location are critical. However, predicting isolated hourly extremes remains a major challenge because small-scale initial errors grow rapidly in numerical weather prediction models. If HEREs are purely stochastic and unconnected, they may be practically unpredictable[16]. Some studies provide a contrasting view, showing that daily extreme rainfall over Northeast and Central India follows an underlying order, occurring preferentially within synoptic systems embedded in the favourable large-scale environment of the active phases of the monsoon intraseasonal oscillation (MISO)[17]. We argue that hourly extremes also follow an order arising from a similar multiscale organization of convection. During active MISO phases, strengthened horizontal winds promote hydrodynamic instabilities and generate synoptic low-pressure systems (LPSs). Rainfall within an LPS occurs predominantly in its southwest quadrant[18,19], implying that HEREs should cluster preferentially in this sector.

Thermodynamic controls may further enhance clustering. Low surface temperatures coincide with low humidity that suppresses HEREs, whereas very high surface temperatures typically occur under high-pressure conditions that also inhibit extreme rainfall. To quantify this organization, we construct a bivariate frequency distribution of HEREs exceeding the 99.9th percentile as a function of surface air temperature and MISO phase. We then extend this framework to a trivariate distribution incorporating surface temperature, total column water vapor (TCWV), and MISO phase. Available hourly rain gauge data are spatially sparse and contain substantial gaps[10], limiting their utility for assessing climate sensitivity and trends. Hourly reanalysis products such as ERA5[20] severely underestimate the intensity of observed HEREs and are therefore unsuitable for this purpose. We rely primarily on satellite observations, using rain gauge data to verify their statistical properties. Satellite products such as CMORPH[21] and IMERG[22] provide high spatial and temporal resolution but span only about 26 years. Although they are not ideal for trend estimation, they are well suited for the clustering analyses pursued here.

Our central finding is a strong multivariate clustering of HEREs, demonstrating that a large fraction of events occur in contiguous clusters exceeding 10,000 km$^2$ within LPSs during the active phases of the MISO and propagate west-north-westward with these systems. This organization reveals that HEREs are not random, isolated bursts but components of coherent storm systems. The order identified here provides a physical basis for medium-range predictability and real-time monitoring of HERE storms. While real-time indices exist for the MISO[23] and for LPSs[19], no comparable framework exists for extreme rainfall events. A key contribution of this study is the development of an algorithm to identify and track HERE clusters. We anticipate that this framework will also be valuable for tracking daily extreme rainfall events.

## 3 Results

### 3.1 Frequency Distribution and Spell Characteristics of Hourly Rainfall data sets

To investigate the predictability of hourly rainfall extremes, the precipitation data must faithfully reproduce the spatiotemporal characteristics of observed extremes. To assess biases in satellite products, we compared the frequency distributions of hourly rainfall from CMORPH, IMERG, and GSMaP-ISRO[24] at the locations of selected IMD rain gauge stations (Fig. 1H). For comparison with reanalysis products, we examined the corresponding distributions from ERA5[20] and

IMDAA[25] at the same station locations (Fig. 1A). ERA5 strongly overestimates very light rain rates below 1.5 mm h$^{-1}$ and underestimates heavier rainfall. More importantly, it severely underrepresents extreme intensities, limiting them to about 25 mm h$^{-1}$, whereas rain gauge observations exceed 110 mm h$^{-1}$. Hourly rainfall near 30 mm h$^{-1}$ in ERA5 is exceedingly rare (Fig. 1A). GSMaP-ISRO and IMDAA also underestimate rainfall above 1.5 mm h$^{-1}$ relative to gauges and fail to capture extremes exceeding 110 mm h$^{-1}$. Among these products, IMDAA and IMERG perform relatively well, reproducing rainfall intensities of up to about 50 mm per hour. In contrast, CMORPH agrees with rain gauge observations up to around 20 mm per hour and slightly underestimates rainfall, although it still captures events approaching nearly 100 mm per hour (Fig.1A). These differences suggest that reanalysis products tend to simulate rainfall over longer periods rather than representing short duration high intensity events. To test this inference, we analysed the rainfall spell duration characteristics across all datasets as a function of intensity, including ERA5, rain gauge observations, and satellite products from CMORPH and IMERG.

We define a light rain spell as a continuous sequence of hours with rainfall between 0.1 and 1.0 mm h$^{-1}$ at a station. To examine low-intensity spells, we analyzed spell-duration distributions for thresholds exceeding 0.1 mm h$^{-1}$ and 1.0 mm h$^{-1}$ (Fig. 1B and 1C). For higher intensities, we examined spell durations exceeding 10 mm h$^{-1}$ (Fig. 1D), 20 mm h$^{-1}$ (Fig. 1E), and 30 mm h$^{-1}$ (Fig. 1F) at the gauge locations (Fig. 1H). All reanalysis products, including ERA5 and GSMaP-ISRO simulate an unrealistically high frequency of light-rain spells persisting for up to 200 hours, whereas gauge observations show that actual spell durations rarely exceed about 30 hours (Fig. 1B and 1C). In contrast, the distribution of daily accumulated rainfall in ERA5 agrees well with that from gauges and satellites (Fig. 1G). Combined with the unrealistic persistence of hourly rainfall, this indicates that daily totals in ERA5 arise from artificially sustained moderate rainfall.

The poor performance of GSMaP-ISRO is unexpected, given its similarity to IMERG. It appears that validation against daily IMD rainfall has smoothed out hourly extremes. IMDAA aligns more closely with satellite and gauge observations, although it overestimates rainfall up to about 1 mm h$^{-1}$. For intensities exceeding 10, 20, or 30 mm h$^{-1}$, IMDAA, like other reanalysis products, fails to capture rare extreme events. Both satellite products overestimate the frequency of spells lasting longer than two hours (Fig. 1F), likely reflecting improved spatial sampling. Despite these biases, satellite precipitation products such as CMORPH provide valuable information on the spatial variability of temperature dependence and the clustering of extremes associated with MISOs. Given its comparable skill in resolving extreme magnitudes, we select CMORPH as the primary dataset for subsequent analyses.

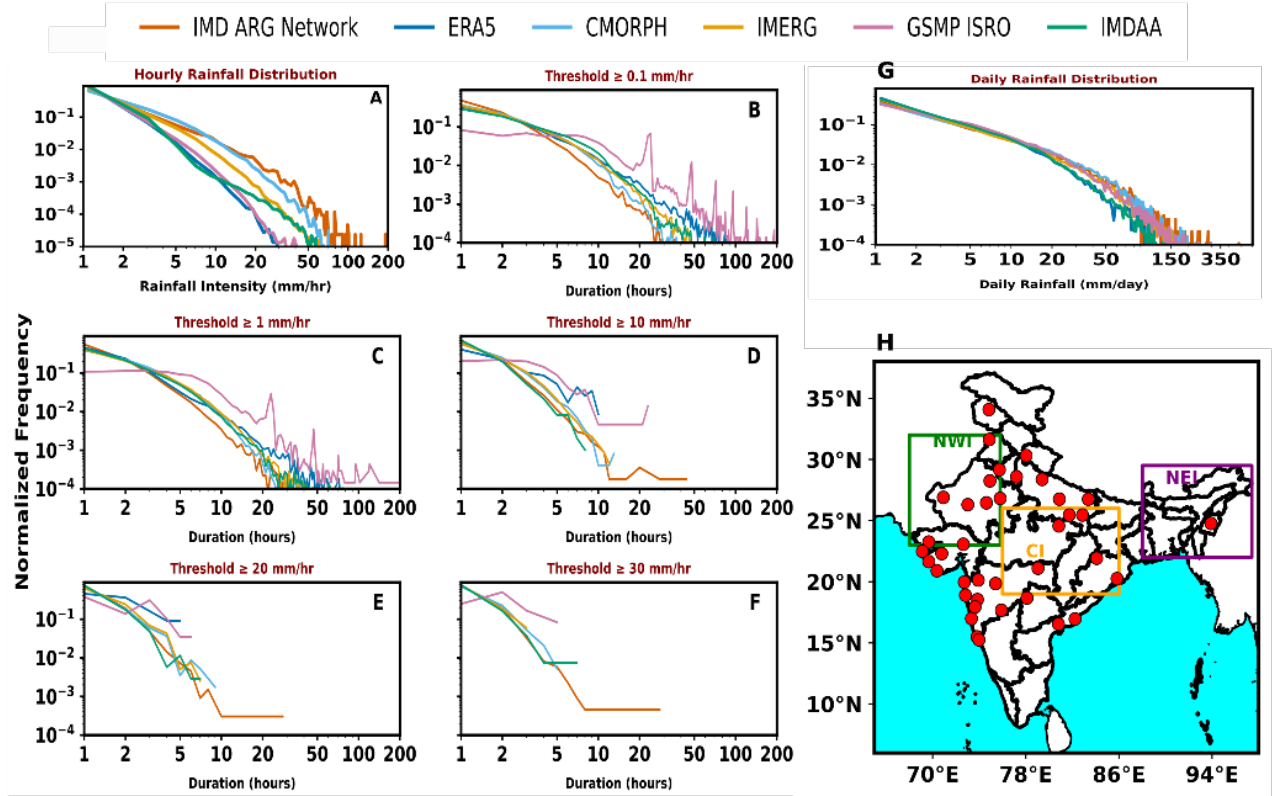

Fig 1 **Spatial distribution and characteristics of rainfall intensity, duration, and observations across datasets over India.** A. Hourly spatial distribution of rainfall intensity and percentage occurrence across different datasets. B.-F. Duration and percentage occurrence of rainfall events exceeding thresholds of 0.1 mm h$^{-1}$, 1 mm h$^{-1}$, 10 mm h$^{-1}$, 20 mm h$^{-1}$, and 30 mm h$^{-1}$, respectively, across the datasets. G. Daily spatial distribution of rainfall intensity and percentage occurrence across different datasets. H. Spatial distribution of IMD hourly rain gauge stations used in this study, shown over India, with boxed areas indicating selected subregions for detailed analysis. The study period spans 1998–2006.

## 3.2 Temperature dependence of hourly extreme rainfall events

Because hourly extremes arise from locally buoyancy-driven instability within the conducive large-scale environment of the MISO or MJO, we examine the dependence of extreme rainfall intensity on daily mean surface air temperature using ERA5 daily temperature and CMORPH daily and hourly rainfall over all India and over Central India, Northeast India, and Northwest India

(Fig. 2). For all three categories of extremes, the 99.0th, 99.5th, and 99.9th percentiles, and for all three regions, the intensity of hourly rainfall extremes increases nonlinearly with temperature. It rises gradually in a low-temperature regime, increases more rapidly in a mid-temperature regime, and then declines sharply in a high-temperature regime (Fig. 2A–D). These three regimes of sensitivity, weak, strong, and decreasing, demonstrate that the temperature–rainfall relationship is nonlinear and reflects both thermodynamic and dynamical controls. Over Central and Northwest India, the low-temperature weak-sensitivity regime is either narrow or absent. Consequently, the weak low-temperature regime seen in the all-India average (Fig. 2A) arises primarily from Northeast India (Fig. 2B).

Over the Indian continent (Fig. 2A), severe hourly extremes exceeding the 99.9th percentile exhibit a sensitivity of 1.41% $K^{-1}$ in the low-temperature regime below 20°C and 4.26% $K^{-1}$ in the mid-temperature regime, close to but lower than Clausius–Clapeyron scaling (Table S1). This behaviour suggests dominance by small-scale, thermodynamically driven moist convection. For moderately severe extremes exceeding the 99.5th percentile, sensitivity increases to 5.72% $K^{-1}$ at mid-temperatures. This enhanced response likely reflects the contribution of mesoscale convective systems, particularly during active MISO phases, when dynamical processes amplify rainfall beyond purely thermodynamic expectations.

The decline in intensity beyond approximately 29°C reflects the tendency for high surface temperatures to occur under clear sky conditions with reduced cloud cover (Fig. S3). This regime corresponds largely to MISO break phases, when organized convection is suppressed by large scale subsidence. The rising portion of the curve therefore represents convectively active low-pressure environments, whereas the declining portion reflects suppressed convection under large-scale high-pressure conditions. A close examination across all regional domains in Fig. 2b to d further reveals a clear-cut oscillatory pattern, with an active convective regime followed by a decreasing regime in which high pressure systems become dominant as cloud cover progressively diminishes. To examine regional contrasts, we extend the analysis to Northeast, Central, and Northwest India (Fig. 1H). Northeast India is distinct because of its humid climate and complex terrain. There, the 99.9th percentile hourly extremes show two dominant regimes, with the decreasing regime confined to a narrow temperature range (Fig. 2B). Hourly extremes increase nearly monotonically at about 4.76% $K^{-1}$ up to approximately 32°C, followed by a brief declining regime extending to about 36°C. Over Central India (Fig. 2C), the 99.9th percentile scaling between 19°C and 26°C reaches approximately 42.26% K$^{-1}$.

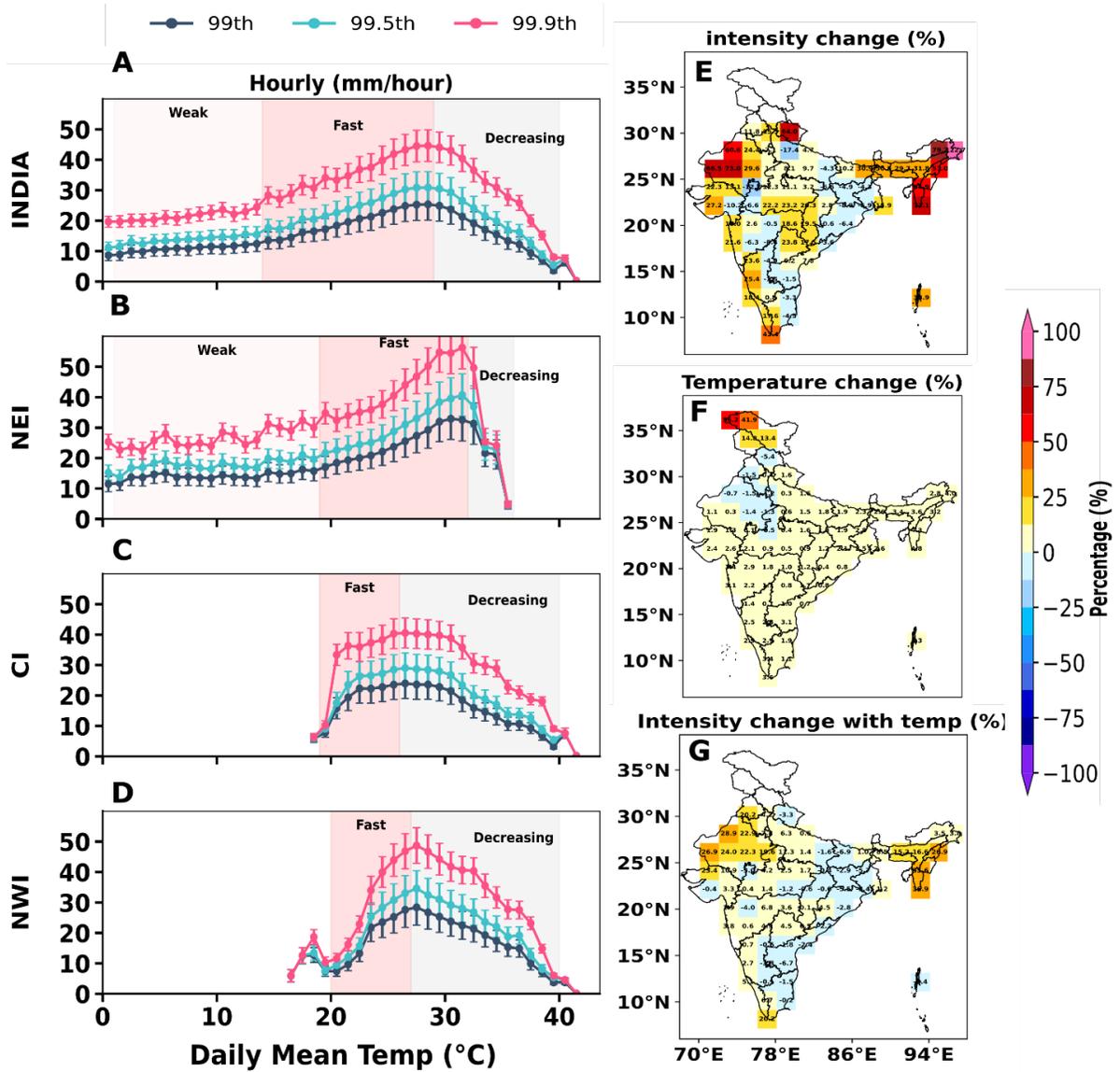

Fig 2 **Temperature dependence and thermal scaling of extreme precipitation over India and its major sub-regions during the summer season.** Temperature dependence of extreme precipitation intensity is shown using CMORPH precipitation and ERA5 2 m temperature data at 0.25° resolution, where the 99th, 99.5th, and 99.9th percentile average precipitation intensities are plotted as a function of daily mean temperature (1 °C bins) for A. All-India, B. Northeast India (NEI), C. Central India (CI), and D. North-West India (NWI). Shaded backgrounds indicate characteristic temperature regimes—Weak, Fast, and Decreasing—highlighting nonlinear shifts in extreme rainfall behavior with temperature. Also shown is the spatial distribution of thermal scaling of summer extreme rainfall at 2° resolution, illustrating E. the percentage change in rainfall intensity during 1998–2023, F. the percentage change in daily mean 2 m air temperature, and G. the percentage change in rainfall intensity with temperature, emphasizing the sensitivity of extreme rainfall to thermal variations over the same period.

In Northwest India, hourly extremes exceeding the 99.9th percentile also display two distinct regimes, with a sensitivity of 43.74% K$^{-1}$ in the moderate-temperature range. This behaviour is consistent with evidence for a westward expansion of the continental ITCZ, which enhances convective instability and dynamical support over the region[26]. Twenty-five years ago, many grid points in this low-rainfall region recorded no extreme events. With the westward expansion of the mean monsoon, summer mean rainfall has increased by up to 200% in parts of the region[26], and hourly extreme intensities have risen proportionately. The spatial distribution of thermal scaling at 2° resolution (Fig. 2D–F) illustrates this contrast. Despite a decreasing trend in mean temperature over Northwest India (Fig. 2F), extreme rainfall intensity has increased by 25 to 60% (Fig. 2E). Over Northeast India, CMORPH shows a substantial rise in thermal scaling, with extreme intensity increasing by 40 to 80% and scaling rates increasing by 25 to 40%.

These contrasting behaviours reflect regional temperature trends. Over Northwest India, mean temperature decreases from 32.23°C to 31.69°C at –0.040°C yr$^{-1}$ (Fig. S1). The westward expansion of the monsoon supplies increased moisture that outweighs the reduced moisture-holding capacity, so lower temperatures coincide with higher extreme rainfall intensity. Over Northeast India, temperature increases from 22.80°C to 23.48°C at +0.033°C yr$^{-1}$ (Fig. S1), a range that lies within the increasing regime in Fig. 2B, so rising temperature corresponds to enhanced extremes. Thermal scaling of daily precipitation extremes shows qualitatively similar behaviour, as discussed in the Supplementary Text with Fig. S2. Together, these results demonstrate that CMORPH reliably captures the temperature sensitivity of hourly and daily extremes over large regions.

### 3.3 Intra-seasonal clustering of extreme hourly rainfall

Because synoptic low-pressure systems tend to cluster during the active phases of the MISO[27,28], and because most extreme rainfall occurs within the conducive environment of these systems, the MISO is expected to organize hourly rainfall extremes as well. Using the MISO index[23], divided into eight phases (Fig. S4), we construct bivariate and trivariate frequency distributions of hourly extreme rainfall events with respect to temperature and total column water vapor over Central India (Fig. 3) and Northeast India (Fig. S5). Figure 2 alone does not reveal whether most HEREs occur within a narrow temperature range. To address this, we first construct a bivariate distribution of hourly extremes over Central India as a function of temperature and MISO phase (Fig. 3A). HEREs occur predominantly within daily mean temperatures of 22 to 32°C, with a sharp concentration between 25 and 28°C, and align closely with MISO phases 5 and 6, corresponding to the active monsoon. This pattern demonstrates that temperature-dependent extremes are strongly modulated by the dominant intraseasonal cycle.

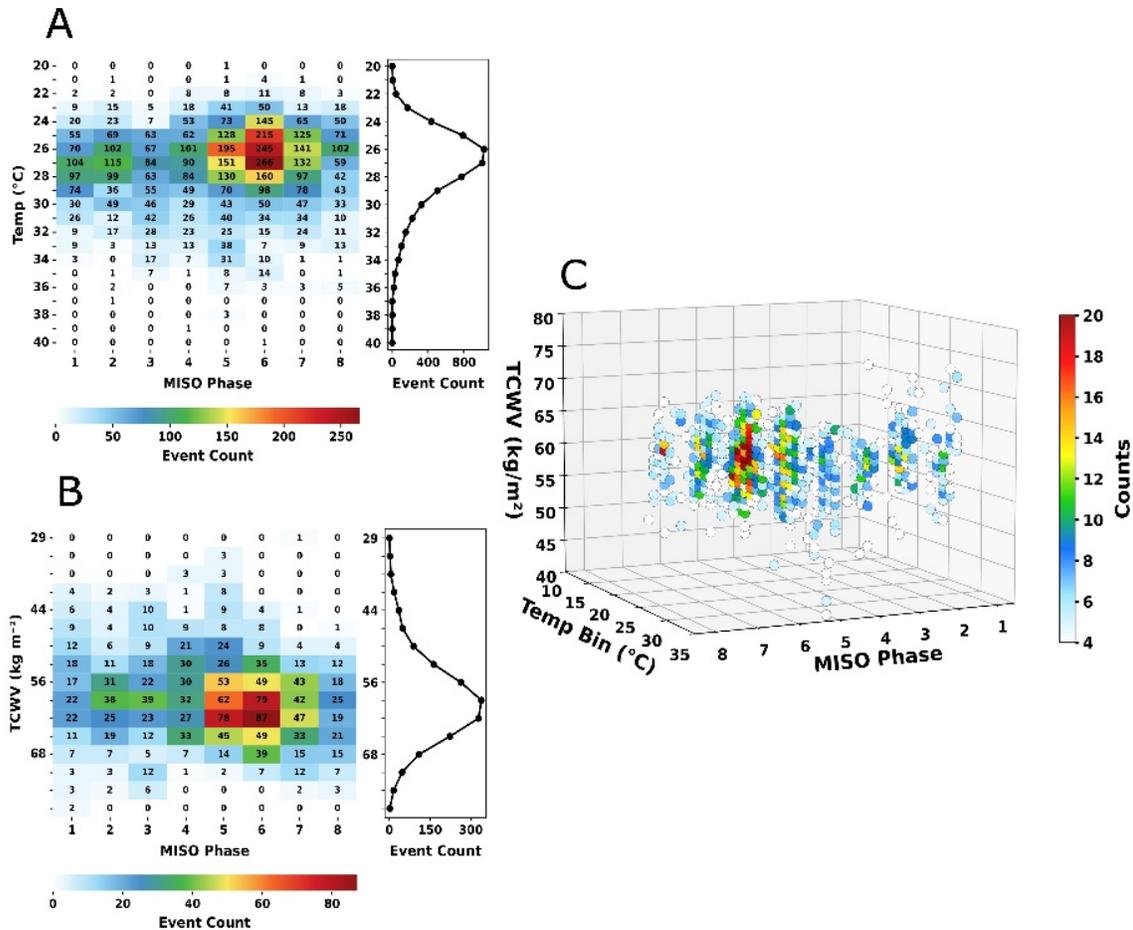

Fig 3 **Bivariate and tri-variate frequency distributions of 99.9th percentile rainfall events over Central India in relation to MISO phase and thermodynamic conditions.** (A) Bivariate distribution of events as a function of MISO phase (x-axis) and daily mean temperature (y-axis), with color shading indicating the event count in each bin. (B) Same as A, but with daily mean total column water vapor (TCWV) on the y-axis. (C) Tri-variate distribution illustrating the joint relationship between MISO phase, daily mean TCWV, and daily mean temperature, where color intensity (or point density) represents the frequency of event occurrences within each three-dimensional bin.

A similar organization emerges in the bivariate distribution of TCWV and MISO phase (Fig. 3B). Over Central India, HEREs are most frequent for TCWV values between 52 and 76 kg m$^{-2}$, with the densest clustering between 54 and 64 kg m$^{-2}$, again during MISO phases 5 and 6. The trivariate distribution combining temperature, TCWV, and MISO phase (Fig. 3C) reinforces this structure. Extreme events form a compact cluster during MISO phases 5 and 6 within a narrow thermodynamic window of 24 to 26°C and 55 to 65 kg m$^{-2}$ TCWV. Local thermodynamic conditions can occasionally persist during break phases, allowing isolated HEREs to occur even outside active periods. This indicates that while the MISO provides the dominant organizing framework, favourable conditions for initiation may extend beyond strictly active phases. The total number of HEREs within the clustered temperature range of 25 to 28°C exceeds those outside this range by more than an order of magnitude (Fig. 3A and B), and the count during MISO phases 5 and 6 is more than twice that of other phases. Together, these features provide direct evidence of potential predictability arising from the combined influence of temperature sensitivity and the MISO.

In Northeast India (Fig. S5), where complex topography shapes the interaction between circulation and moisture, the MISO signature remains evident but manifests differently. HEREs span a broader temperature range of 9 to 30°C, with most clustered between 25 and 29°C, and occur for TCWV values between 44 and 68 kg m$^{-2}$. Rather than concentrating within a single pair of MISO phases, extremes appear throughout the intraseasonal cycle. This behaviour reflects how MISO-driven moisture transport and wind anomalies interact with orography, enabling extreme rainfall across a wider range of phases.

### 3.4 Storm Characteristics Across MISO Phases

To further examine how HEREs organize within low-pressure systems under intraseasonal modulation, we classify extreme rainfall events into HERE storm systems spanning spatial scales from 100 to 300 km. Each HERE storm is defined as a contiguous region exceeding the 99.9th percentile of hourly rainfall, representing a coherent cluster of extremes. This definition isolates the most intense core of each system, ensuring that the reported storm diameter corresponds to the region with the highest probability of HEREs rather than to the broader rain field. Figure 4 shows the spatial distribution and tracks of 100 km storm cores during 2022, overlaid on climatological MISO phase composites for 1998 to 2023. Storms preferentially initiate during active MISO phases and propagate northward as the intraseasonal cycle evolves. Even during break conditions over the Indian landmass in Phases 1 and 2, the MISO remains active over adjacent oceanic regions to the south, where HERE storms form and subsequently migrate northward toward the subcontinent. The clustering and systematic migration of these storms across MISO phases demonstrate that storm genesis, movement, and extreme rainfall are intrinsically governed by the evolving large-scale intraseasonal environment. Extending the analysis to larger systems of 150km and 200km (Fig. S6) reveals an even tighter alignment with specific MISO phases, indicating that larger storm systems are more strongly constrained by the MISO.

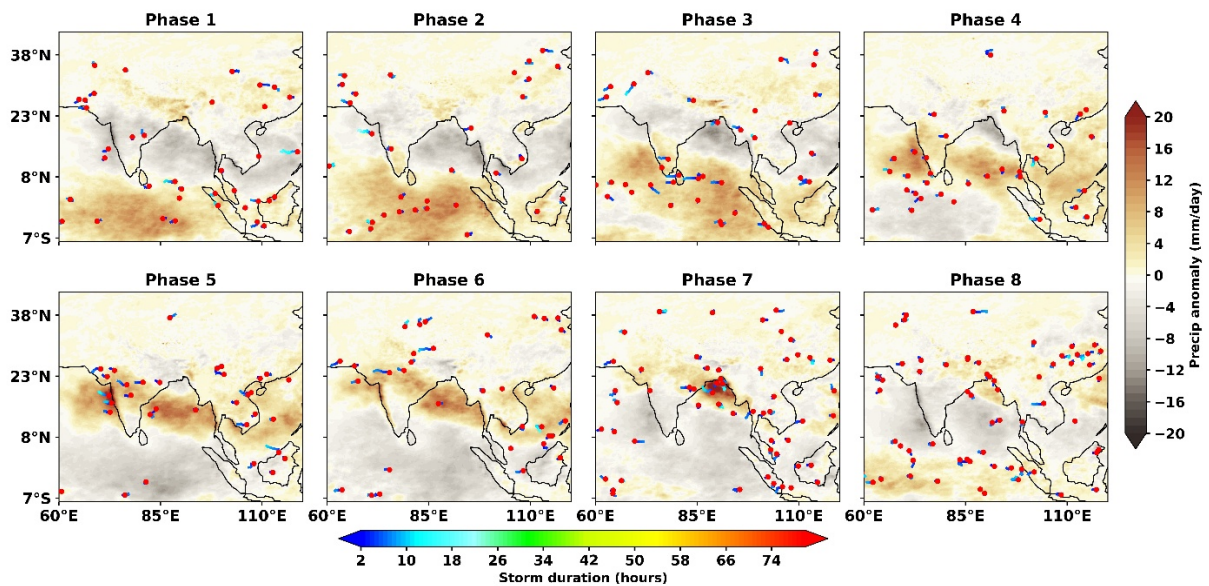

Fig 4 Tracks of 100km storm cores generated in 2022, overlaid on MISO phases derived from the 1998–2023 climatological cycle.

Within the MISO environment, HERE storms are further organized by LPSs. To examine this structure, we generate LPS tracks over the Indian summer monsoon region for 1998 to 2023. Most LPSs originate within the monsoon trough (Fig. 5A). Figure 5B presents a composite of all LPS tracks, constructed at 6-hour intervals by aligning system centers at their maximum footprint. Consistent with earlier studies showing that rainfall preferentially occurs in the southwest quadrant of an LPS[18,27], HERE storms cluster predominantly in this sector (Fig. 5B). The northwest and southwest quadrants contribute 32.3% and 40.9% of events, respectively, within the most concentrated region located between relative latitudes 0 to −5 and longitudes −5 to 5 from the LPS center. This region coincides with composite stream function values of −0.833 to −1.500, indicating enhanced moisture transport. Although storms may initiate in different sectors, strong flow and moisture convergence enable them to reach their maximum footprint in the southwest quadrant, where the highest counts of HEREs, reaching 18 to 20, are observed.

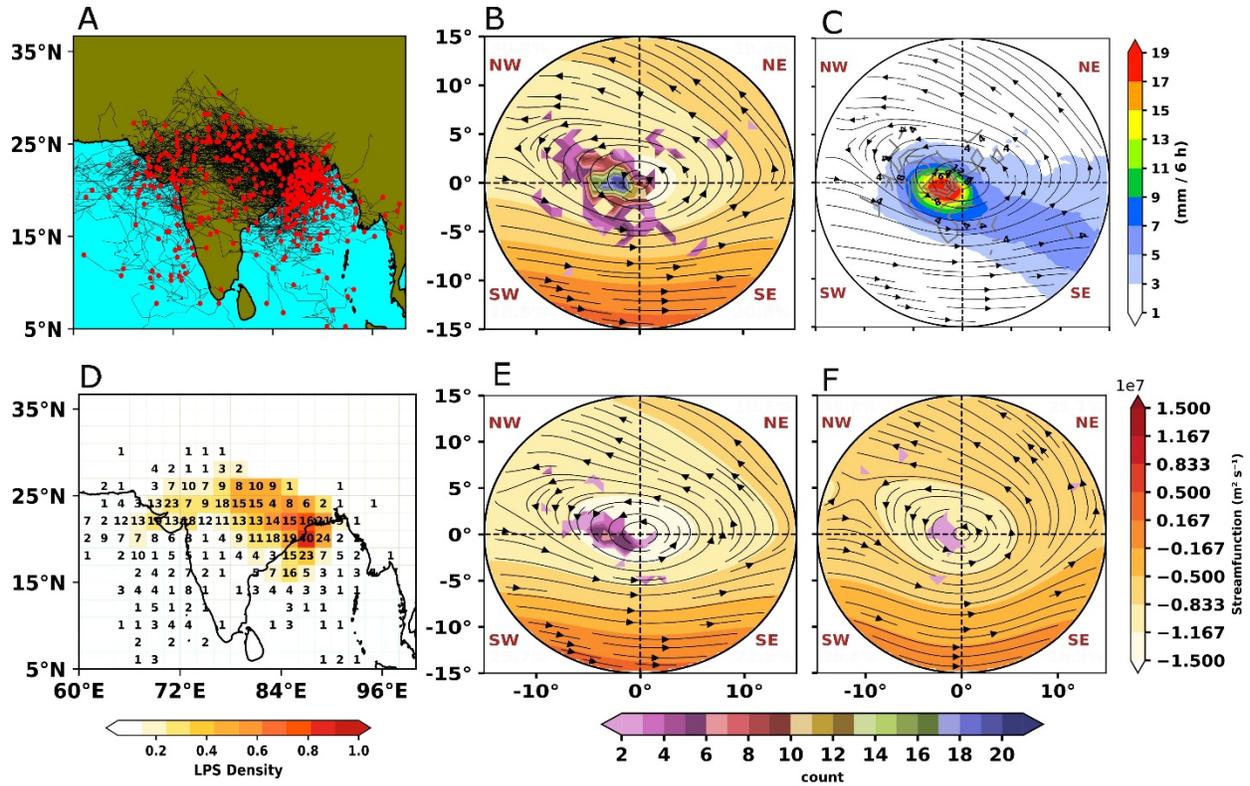

Fig 5. **LPS tracks and ERE storm density relative to the LPS center under different MISO phases.** A. Tracks of all low pressure systems (LPSs) generated during 1998 to 2023 over the Indian core monsoon region. B. Composite of all LPS six hourly center points obtained during the analysis, overlaid with the spatial density of 100 km ERE storms within a 15° radius centered on the LPS. C. Same as (B) but overlaid with six hourly accumulated rainfall from CMORPH at 0.25° resolution. D. Frequency density map of LPSs and HERES over the monsoon core region. The color bar indicates the normalized density of LPS center points at 2° resolution. Higher intensity denotes a larger number of LPS centers passing through a given grid cell. The numbers indicate the total HERES associated with LPSs during 1998 to 2023. E. Same as (B) but for the MISO active phase over land, primarily over central India. F. Same as (B), but for the MISO break phase.

Analysis of larger systems exceeding 100 km (Fig. S8) shows that storm cores increasingly cluster closer to the LPS center, where circulation is strongest. The fraction of storms in the southwest sector rises monotonically with system size (Table 1), and within a 5° radius, 300 km systems occupy 75% of occurrences. Hourly extremes defined by the 99.5th percentile display similar behaviour, with the fraction in the southwest quadrant increasing with size and reaching 58.7% for 300 km systems. To assess intraseasonal modulation, we separate LPS tracks by active and break MISO phases over Central India. During active phases, LPSs are stronger, with stream function values of −1.167 to −1.500, compared with −0.5 to −0.833 during break phases. Consistent with this enhanced circulation, 100 km HERE storms cluster more densely during active phases, with counts of 4 to 6, whereas such events are rare during break phases. Storm motion reflects the superposition of LPS and MISO wind anomalies. Mapping the normalized frequency of LPS passage on a 2° grid and overlaying HERE counts (Fig. 5D) reveals that HEREs are most frequent where LPS density is highest, demonstrating a strong spatial correspondence between LPS occurrence and the clustering of hourly extremes.

Table 1. Percentage occurrence of HERES of different sizes across the four quadrants of LPS with 5-degree radius

| Storm Size | 99.9 | | | | 99.5 | | | |
|---|---|---|---|---|---|---|---|---|
| | NE | NW | SW | SE | NE | NW | SW | SE |
| 100km | 12.9% | 32.3% | 40.9% | 14.0% | 12.3% | 32.6% | 35.9% | 19.2% |
| 150km | 5.6% | 41.7% | 41.7% | 11.1% | 9.3% | 32.2% | 40.2% | 18.3% |
| 200km | 9.8% | 43.9% | 46.3% | 0% | 8.9% | 27.6% | 52.1% | 11.5% |
| 250km | 5.9% | 11.8% | 82.4% | 0% | 5.1% | 38.0% | 44.3% | 12.7% |
| 300km | 0.0% | 25.0% | 75.0% | 0% | 6.5% | 32.6% | 58.7% | 2.2% |

To characterize storm properties, we focus on the monsoon core zone spanning 18 to 28°N and 65 to 88°E. Storms derived from the 99.9th percentile exhibit strong intraseasonal modulation in both intensity and duration(Fig. S7). During active MISO phases 5 and 6, 100 km systems reach intensities of 125 to 150 mm h$^{-1}$, whereas during break phases they remain near 50 to 70 mm h$^{-1}$. Larger systems occur mainly during active phases, and HERE counts increase from 60 to 78 during break phases to 155 to 213 during active phases. Over 1998 to 2023 ,storm occurrence peaks sharply during MISO phases 5 and 6. Across all phases, 100 km systems are the most frequent, while systems larger than 100 km become progressively more common during active phases. Storm duration also depends on MISO state, with large systems persisting for 40 to 80 hours during active phases but only 20 to 30 hours during break phases. This enhanced growth and longevity highlight the role of large-scale convective and dynamical forcing in the MISO environment. The largest systems, dominant during active phases, exhibit the most intense cores, identifying them as primary contributors to regional extreme rainfall and providing a physical basis and methodology for predicting the most severe HEREs, including cloudburst-type events.

## 4 Discussion

Extreme rainfall intensity across India exhibits strong nonlinear scaling with temperature, shaped by both thermodynamic and dynamic processes. At lower temperatures, extremes broadly follow Clausius–Clapeyron expectations, whereas several regions display super-CC scaling at higher temperatures, reflecting enhanced convective organization and large-scale moisture convergence. The decline in extreme intensity beyond a critical temperature occurs during MISO break phases, when large-scale subsidence produces clear-sky conditions and elevated surface temperatures. Building on this behaviour, the most novel result of this study is the trivariate clustered relationship among extreme rainfall, temperature, MISO phase, and column-integrated water vapor, revealing a multiscale interaction that governs the occurrence of extreme events[17] . These findings demonstrate that most hourly rainfall extremes are not isolated, random bursts of convection but arise preferentially within specific thermodynamic states that occur during particular phases of the MISO. The spatial and temporal coherence of extremes within narrow ranges of temperature, moisture, and intraseasonal phase shows that their predictability is fundamentally controlled by the large-scale monsoon system rather than by local triggering alone. This provides a mechanistic basis for anticipating periods of heightened extreme rainfall risk before individual convective cells develop, extending actionable lead time beyond the limits of conventional weather models.

For hourly extremes with lifetimes of 3 to 6 hours, slow modulation by the MISO on 30 to 60 day time scales supplies the memory required for predictability, while embedded low-pressure systems further cluster events into coherent HERE storms. Our framework tracks HERE storms within individually tracked LPSs under a given, real-time monitored MISO phase. Because state-of-the-art coupled ocean–atmosphere models can predict MISO phase evolution more than two weeks in advance[29,30] , prediction of HERE storms may be feasible more than a week ahead. Present dynamical weather models cannot realize this potential because of rapid error growth, and linear statistical models fail because of the inherent nonlinearity of multiscale interactions. Recent advances in deep learning demonstrate that large-scale weather prediction skill can exceed that of dynamical models[31,32]. By providing a physics-based road map for training such models, this study opens a path toward extended-range prediction of the most destructive HERE storms.

These conclusions must be interpreted considering important limitations. Satellite datasets, despite their spatial completeness, underestimate or oversimplify the upper tail of hourly rainfall intensity distributions, and the rain gauge network remains sparse, particularly in regions of complex terrain. These constraints introduce uncertainty in the precise magnitude of sensitivities and in the extremity of upper-percentile events.

**Materials and Methods**

**Datasets**

This study examines the characteristics and sub-seasonal modulation of extreme hourly rainfall across India during the summer monsoon using rain gauge observations, satellite-derived precipitation products, and atmospheric reanalyses. Hourly rain gauge observations from the India Meteorological Department were available for 2000 to 2006[10] and serve as the benchmark for evaluating remotely sensed rainfall over the Indian subcontinent.

Satellite precipitation estimates were analyzed for 1998 to 2023 from three independent products: the Climate Prediction Center Morphing Technique CMORPH dataset at 8 km resolution[21], the Integrated Multi-satellite Retrievals for the Global Precipitation Measurement IMERG final-run product at 10 km resolution[22], and the Global Satellite Mapping of Precipitation ISRO GSMaP-ISRO dataset at 0.1° resolution[24]. These products enable detailed characterization of the spatial organization, intensity distribution, and upper-tail behaviour of extreme rainfall at sub-daily time scales across diverse rainfall regimes in India.

To characterize the thermodynamic environment of extreme precipitation, we use variables from the European Centre for Medium-Range Weather Forecasts fifth-generation reanalysis ERA5[20]. All ERA5 fields are taken at hourly resolution, including 2 m air temperature, precipitation, and total column water vapor. The 2 m temperature represents near-surface thermal conditions relevant to convective instability, while total column water vapor quantifies vertically integrated moisture, enabling a joint assessment of thermal and moisture controls on extremes. To further evaluate rainfall intensity and spell characteristics, we also use hourly precipitation from the Indian Monsoon Data Assimilation and Analysis reanalysis IMDAA[25], which provides enhanced spatial resolution and dynamical consistency for the Indian region. For identification and tracking of low-pressure systems, we use ERA5 zonal and meridional winds and relative humidity at 850 hPa, together with geopotential height, 10 m winds, and the land–sea mask.

**Frequency Distributions and spell Characteristics**

To evaluate how well satellite and reanalysis products represent sub-daily rainfall characteristics, we performed a unified analysis of hourly rainfall intensity distributions and rainfall spell durations at IMD rain gauge locations (Fig. 1G). Hourly precipitation from each dataset was collocated with gauge stations and grouped into fixed 2 mm h$^{-1}$ intensity bins. Counts within each bin were normalized by bin width and by the total number of events to obtain probability density functions. To analyze rainfall spells, hourly precipitation was thresholded at five discrete levels, 0.1, 1, 10, 20, and 30 mm h$^{-1}$, and converted into binary rain and no-rain sequences. Contiguous rainy hours were identified as individual spells. Spell duration was computed in hours, and spell-duration frequencies were normalized to produce duration–probability distributions. A schematic illustration of this procedure is provided in Fig. S9.

**Realtime MISO Monitoring**

To construct the MISO index[23], we computed daily rainfall anomalies from CMORPH precipitation at 0.25° resolution for 1998 to 2023 by removing the climatological annual cycle, defined using the mean and the first three harmonics at each grid point. To isolate intraseasonal variability associated with the Monsoon Intraseasonal Oscillation, anomalies were longitudinally averaged over 60.5°E to 95.5°E and analyzed over 12.5°S to 30.5°N during the JJAS season. We applied extended empirical orthogonal function analysis using a 15 day lag window with 1 day increments to capture the dominant 24 to 40 day variability. The leading two EEOF modes, retained based on sampling error estimates, form a coherent propagating pair that represents the full life cycle of northward-moving rainfall anomalies. Their corresponding principal components define the MISO1 and MISO2 indices (Fig. S4I).

We characterize the daily evolution of the MISO in a two-dimensional phase space defined by MISO1 and MISO2 and divide it into eight phases representing the systematic progression of convection over the Indian monsoon region (Fig. S4J–L). For 1998 to 2023, we assign each day to a MISO phase and use these dates to generate composite rainfall anomalies (Fig. S4A–H).

**ERE Size Clustering and tracking**

Extreme rainfall systems were identified using an object-based clustering framework applied to hourly precipitation fields. For each year during the JJAS season, grid cells exceeding the annual 99.5$^{th}$ or 99.9$^{th}$ percentile precipitation threshold were first identified and converted into a binary exceedance mask. This thresholding isolates intense rainfall cores while minimizing the influence of widespread stratiform precipitation, thereby capturing dynamically active storm regions associated with extremes.

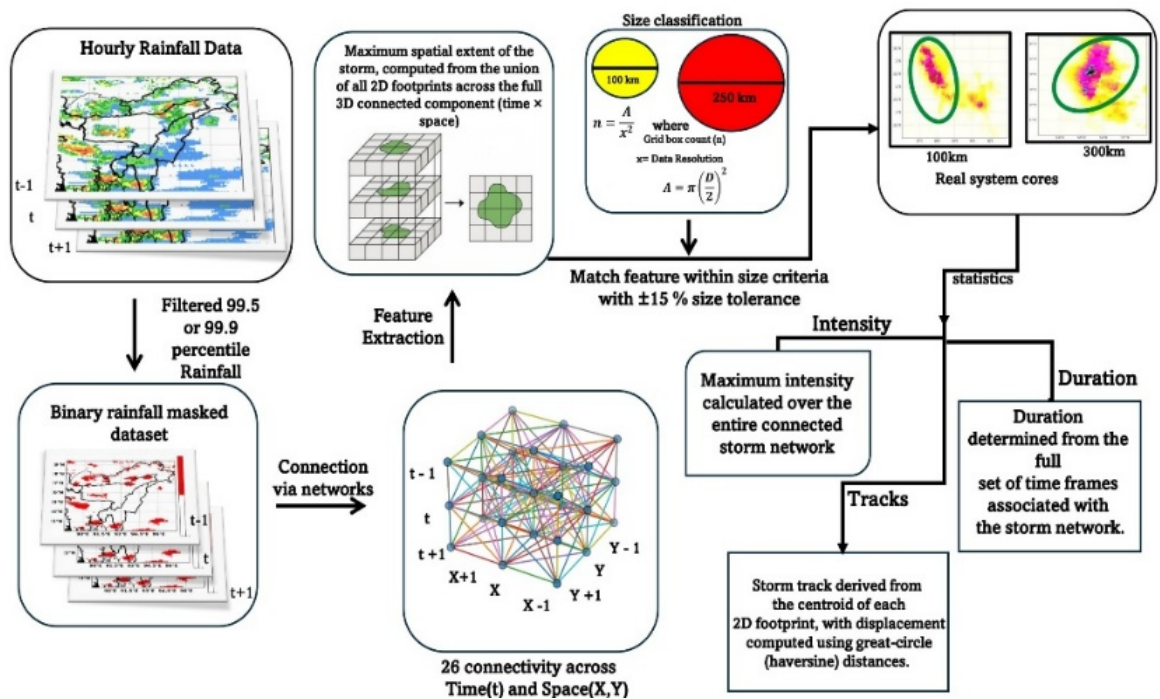

Fig 6 Schematic represents of the algorithmic workflow to cluster and tracks ERE storms.

Spatiotemporally coherent extreme rainfall systems were then extracted using three-dimensional connected-component labelling across latitude, longitude, and time. In this framework, any voxel-connected structure is treated as a single evolving storm entity, allowing systems to be tracked continuously from initiation through maturity and dissipation[33,34]. Voxel connectivity naturally breaks when a storm core weakens or fragments, ensuring physically meaningful termination of storm lifecycles. For each detected cluster, we quantified its horizontal footprint at every hourly time step and defined the characteristic size of the event as the maximum number of extreme grid cells attained during its lifetime. These footprints were converted into equivalent spatial scales and assigned to predefined size categories spanning 50 to 300 km, with a tolerance of ±15%, to stratify events by physical scale and enable consistent interannual comparison. Events falling outside these ranges were excluded to maintain a uniform classification framework.

To characterize storm motion and evolution, we computed hourly centroid locations for each cluster from the mean latitude and longitude of all extreme grid cells at a given time. The resulting centroid time series was used to reconstruct storm trajectories, and straight-line displacement was estimated using great-circle geodesic calculations that account for Earth's curvature[35]. Rainfall intensity for each system was defined as the maximum precipitation rate observed across all connected voxels during the storm's lifetime. A schematic overview of the clustering methodology is shown in Fig. 6.

**Monsoon LPS (Low pressure System) Tracking and clustering**

For low-pressure system (LPS) tracking, we employed the automated Lagrangian pointwise feature tracking algorithm Tempest Extremes[36–38] to extract LPS track information from the reanalysis data. The tracking framework is based on a MapReduce approach implemented in two stages: (i) parallel identification of candidate features at each time step using objective thresholds and criteria that ensure the presence of a closed circulation, and (ii) temporal linking of nearby candidates across successive time steps to construct continuous system tracks, while filtering out transient or non-propagating features. Stream function at 850 hPa was used as the primary diagnostic variable[38] for LPS detection. The tracking was performed over the Indian Summer Monsoon Region (ISMR) during the JJAS season for the period 1998–2023, within the spatial domain 7°S–42°N and 55°E–110°E. After obtaining the LPS tracks, each LPS was followed individually at 6-hourly intervals. For every LPS, the system centre was identified at each 6-hourly time step, and the track was continued as long as the system persisted. The identified LPS centre at each time step was treated as the reference point for further analysis. Around each 6-hourly LPS centre, a circular domain with a radius of 15° was constructed. Within this radius, storm systems with spatial sizes ranging from 150 to 300 km were identified. For each storm, only the location corresponding to the time at which it attained its maximum spatial footprint was retained, ensuring that each storm was counted once and represented at its most developed stage. This procedure was repeated for all LPS tracks over the 25-year study period (1998–2023). Using the 6-hourly LPS centers as reference points, storm occurrences were composited in a Lagrangian framework centered on the LPS position. To quantify storm density, the 15° radius composite domain was divided into 1° × 1° latitude–longitude grid boxes, and the number of storm events within each grid box was counted. The resulting gridded counts represent the spatial density of storm occurrences relative to the LPS centre.

**Temperature Dependence and Multivariate Clustering of ERE**

To quantify the temperature dependence of extreme precipitation, hourly precipitation $P(day, hour, lat, lon)$ and daily mean 2m air temperature $T_{2m}(day, lat, lon)$ were analysed during the summer season (April–September), retaining only days $D$ with 24-hour precipitation records and collocated daily temperature fields. Hourly precipitation was grouped by calendar day to form daily blocks. Short-duration precipitation extremes were defined for each grid point $(lat, lon)$ as

$$P^{(3)}(D, lat, lon) = (\text{Top }_3\{P(D, hour, lat, lon)\}_{hour=1}^{24}),$$

with values <0.1mm excluded.

Daily mean temperature was discretized into 1 °C bins $[\theta_j, \theta_{j+1})$. For each temperature bin $j$, precipitation samples were defined as

$$S_j^{(3)} = \{P^{(3)}(D, lat, lon) \mid T_{2m}(D, lat, lon) \in [\theta_j, \theta_{j+1})\}.$$

Bin-specific extreme thresholds were computed using upper-tail percentiles,

$$\tau_{z,j}^{(3)} = \text{Percentile }_z \left(S_j^{(3)}\right), z \in \{99, 99.5, 99.9\}.$$

To estimate spatial scaling rates, the three highest hourly precipitation values were first extracted at each grid point for every day. All selected hourly extremes were then pooled within each calendar year, and an annual extreme precipitation index was computed as the 99.5th percentile of these values. Long-term changes in extreme precipitation were assessed using grid-wise linear regression, yielding both absolute and relative percentage trends over the analysis period. Summer 2-m air temperature trends were calculated in parallel using annual mean summer temperature at each grid point. To quantify temperature dependence, the annual extreme precipitation indices were regressed against collocated annual mean summer 2-m air temperature, providing spatially resolved precipitation–temperature scaling rates.

To examine joint thermodynamic and dynamical controls, Total column water vapour (TCWV) and daily mean 2m temperature ($T_{2m}$) are combined with The Monsoon intra-seasonal oscillation (MISO) phases to construct a trivariate phase space. Daily TCWV and $T_{2m}$ are discretized into bins $[\omega_i, \omega_{i+1})$ and $[\theta_j, \theta_{j+1})$, respectively, and each day $D$ is assigned a MISO phase $M(D) = m$. For each joint bin $(i, j, m)$ short-duration precipitation extremes P, defined by the maximum of the three largest hourly precipitation values, are sampled as

$$S^3_{i,j,m} = \{P^3(D, lat, lon) | tcwv(D, lat, lon) \in [\omega_i, \omega_{i+1}), T_{2m}(D, lat, lon) \in [\theta_j, \theta_{j+1}), M(D) = m\}.$$

Extreme thresholds are then defined using 99.5 percentiles as

$$\tau^{(3)}_{z,i,j,m} = percentile_{99.9}(S^3_{i,j,m})$$

Bi-variate analyses are treated as lower dimensional projections of this framework, obtained by conditioning precipitation extremes on either thermodynamic variable alone (TCWV or $T_{2m}$) combined with MISO phase. Counts of extreme events were obtained by identifying days on which precipitation exceeded the corresponding percentile thresholds within each phase-space bin.

**Acknowledgments**
BNG is grateful to Gauhati University, Guwahati for support through ANRF Prime Minister Professor. BKB acknowledges the Department of Instrumentation and USIC and ST radar Centre for the Project Associate Program. BKB and US acknowledge Gauhati University for providing support to carry out the research work. The Indian institute of Tropical meteorology is funded by Ministry of Earth Sciences, Govt. of India. DS acknowledges the funding from IndusInd Bank through the CSR Project (CR23242519AEINIB002696). S. Das is grateful to the Department of Science and Technology (DST), India for the INSPIRE fellowship with grant No. DST/INSPIREFellowship/2020/IF20012

**Funding:** This research did not receive any grant from funding agencies in the public, commercial, or not-for-profit sectors

**Author contributions:**
Conceptualization: BNG, BKB
Methodology: BKB
Investigation: BNG, BKB, DS, MK, US, SD
Visualization: BNG, BKB, MK, US
Supervision: BNG, MK, US
Writing—original draft: BNG, BKB
Writing—review & editing: BNG, BKB, DS, MK, SD

**Competing interests:** The Authors declare that they have no competing interests.

**Data and materials availability:** Data related to this paper can be downloaded from: CMORPH https://www.ncei.noaa.gov/products/climate-data-records/precipitation-cmorph, IMERG https://www.earthdata.nasa.gov/dashboard/data-catalog/GPM_3IMERGDF.v07, GSMap-ISRO https://www.mosdac.gov.in/gsmap-isro-rain, ERA5 https://cds.climate.copernicus.eu/datasets/, IMDAA https://nwp.ncmrwf.gov.in/reanalysis, IMD Hourly ARG data available upon reasonable request.

Supplementary Materials for

# Sub-seasonal Modulation and Predictability of Indian monsoon hourly Rainfall Extremes

Bijit Kumar Banerjee[1,2], Devabrat Sharma[3,4], Mahen Konwar[5], Simanta Das[6], Utpal Sarma[1,2] and B. N. Goswami[2*]

[1]Department of instrumentation and USIC, Gauhati University, Guwahati-781014, Assam, India
[2]ST radar Centre, Gauhati University, Guwahati-781014, Assam, India
[3]Department of Aerospace Engineering, Indian Institute of Technology Madras, Chennai-600036, India
[4]Centre of Excellence for Studying Critical Transitions in Complex Systems, Indian Institute of Technology Madras, Chennai-600036, India
[5]Indian Institute of Tropical Meteorology, Pune-411008, India
[6]Department of Physics, Cotton University, ,Guwahati-781001, Assam, India

*Corresponding author Email: bhupengoswami100@gmail.com

**This PDF file includes:**

Supplementary Text
Figs. S1 to S9
Tables S1

**Supplementary Text**

**What is known about temperature dependence and drivers of hourly extremes elsewhere?**

The sub-daily rainfall extremes on hourly or sub-hourly time scales are associated with deep convection[39]. While large-scale circulation and thermodynamics maybe responsible for initiation, local convective potential, diurnal cycle, orography and land-surface conditions can influence its intensification. Therefore, the frequency and intensity of hourly extremes and their trends may vary from place to place. Using a quality-controlled database of hourly records sampling different climate regions including the United States, Australia, the British Isles, Japan, India and peninsular Malaysia over the 1950–2016 period[40], examined *climatologica*l features of observed annual maximum precipitation on hourly and daily timescale find that most yearly 1-hr maximum precipitation events are embedded in short duration storms of 1-5 hr duration while most daily maximum precipitation events is from accumulated precipitation from storms lasting for <15 hrs. Based on precipitation data of 5-min to 12h duration from 100 meteorological stations in Canada[41], find that the intensity- temperature relationship is sensitive to duration of the event and the local climate. Our objective here is to estimate the change in frequency and intensity of hourly extremes over India and their temperature dependence. While some studies using hourly rain gauge data and regional climate model simulations)[5,39,42] investigate the temperature dependence of hourly rainfall extremes and find it to increase at twice as compared to Clausius Clapeyron in extratropical locations in Europe, over the Indian monsoon region in much warmer temperatures and conditionally unstable atmosphere, the relationship could be significantly different. Using sub-daily precipitation data at 23 urban stations supported by other sub-daily precipitation products like TRMM and COMRPH[43], examine the contribution of dynamic and thermodynamic factors on the temperature dependence of sub-daily rainfall extremes. They find that the half-hourly precipitation extremes show higher contributions from both thermodynamic (~10%/K) and dynamic (~15%/K) scaling than daily (6%/K and 9%/K, respectively) extremes. As the Northeast India (NEI) is a region very intense convective activity aided by interaction of winds with the meso-scale orography leading to daily rainfall extremes are increasing at 12.5%/K[44], we anticipate the hourly extremes to increase even faster. This region is disproportionately battered by the increasing trends of frequency and intensity of rainfall extremes as they occur on a three-fold larger mean rainfall compared to that over the Central India. In the absence of a dense network of rain gauge stations with hourly rainfall records, the only way we could get estimates of temperature sensitivity of hourly rainfall extremes and their spatial inhomogeneity over India is using satellite products. One of the limitations of high temporal and spatial resolution satellite products like CMORPH[21] and IMERG[22] is the relatively short length of data (1998- 2023). These satellite-based precipitation products a high temporal and spatial resolutions has accelerated their use in various hydro-meteorological and hydro-climatological applications. Validation studies of CMORPH and IMERG with respect to in-situ rain gauge stations[45–47] indicate that the biases of CMORPH and IMERG in detecting the frequency and intensity of different rain rates depend on local terrain, urbanization[48] and other land surface conditions. This implies potential for significant artificial spatial variability in trend of frequency and intensity of extreme events. In the present study, we examined the frequency and duration of spells of extreme hourly rain rates exceeding 10 mm/hr (Figure 1g) over 39 locations of hourly rain gauge stations and find that the satellite products detect the short duration (1-3 hrs) intense hourly rainfall well while detects more extreme rainfall events lasting 3-10 hrs compared to rain gauge stations. It may be noted that even the hourly rain gauge stations have uncertainty arising from sampling issues. Therefore, notwithstanding these intrinsic limitations of the satellite products, especially for examining spatiotemporal variations of hourly extreme rainfall on a relatively large spatial scale, CMORPH and IMERG may be useful. We need also to remember that the Indian monsoon has a multi-decadal variability with a period of about 60-years[49]. In that context, another limitation of the satellite products of rainfall is that they are available for a short recent period namely between 1998 and 2023. The trends and sensitivity of extreme rainfall events derived from these products could be biased by the phase of the multi-decadal oscillation during this period. The results would be interpreted keeping this factor in mind.

**Temperature Dependence of ERE on daily scales**

Across India, daily rainfall extremes exhibit a clear nonlinear dependence on daily mean temperature. The intensity increases slowly at lower temperatures, strengthens rapidly beyond ~20 °C, and weakens at temperatures higher than ~27 °C, indicating distinct weak, strong, and decreasing sensitivity regimes. For all-India daily extremes exceeding the 99.9th percentile, the high-temperature sensitivity reaches 5.00 % K$^{-1}$, while for the 99.5th percentile it is 7.8 % K$^{-1}$, suggesting the role of mesoscale–synoptic interactions in enhancing moisture convergence over large spatial scales. Over NEI, the 99.9th percentile daily extremes show two temperature regimes, with sensitivities of 4.34 % $K^{-1}$ in the weak regime and 5.57 % $K^{-1}$ in the strong regime, reflecting the influence of complex topography and humid conditions. In CI, daily extremes display a sharp transition near 20 °C from weak to strong dependence, with a high-temperature sensitivity of 44.19% $K^{-1}$, indicating a rapid intensification of extremes under favorable thermodynamic and dynamical conditions. Over NWI, daily extremes exceeding the 99.9th percentile also show two distinct regimes, with an exceptionally strong sensitivity of 110.39 % $K^{-1}$ in the high-temperature regime, consistent with enhanced convective instability and large-scale dynamical support over the region.

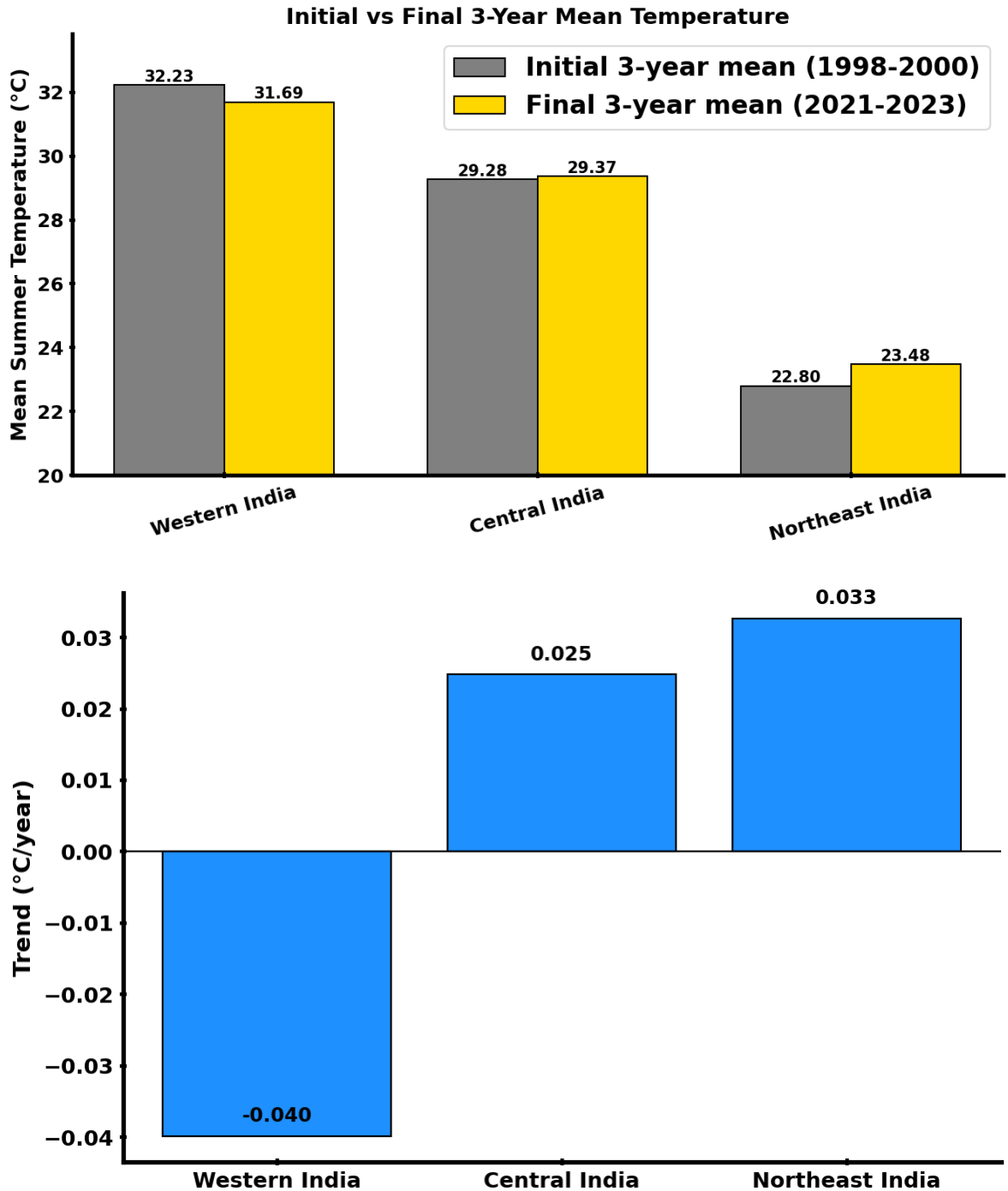

**Fig. S1.**

Upper panel: Comparison of the initial (first 3-year mean) and final (last 3-year mean) summer (April–September) near-surface air temperatures over different subregions of India. Lower panel: Spatially averaged summer temperature trends over the same regions, calculated using only grid points with statistically significant trends ($p < 0.05$).

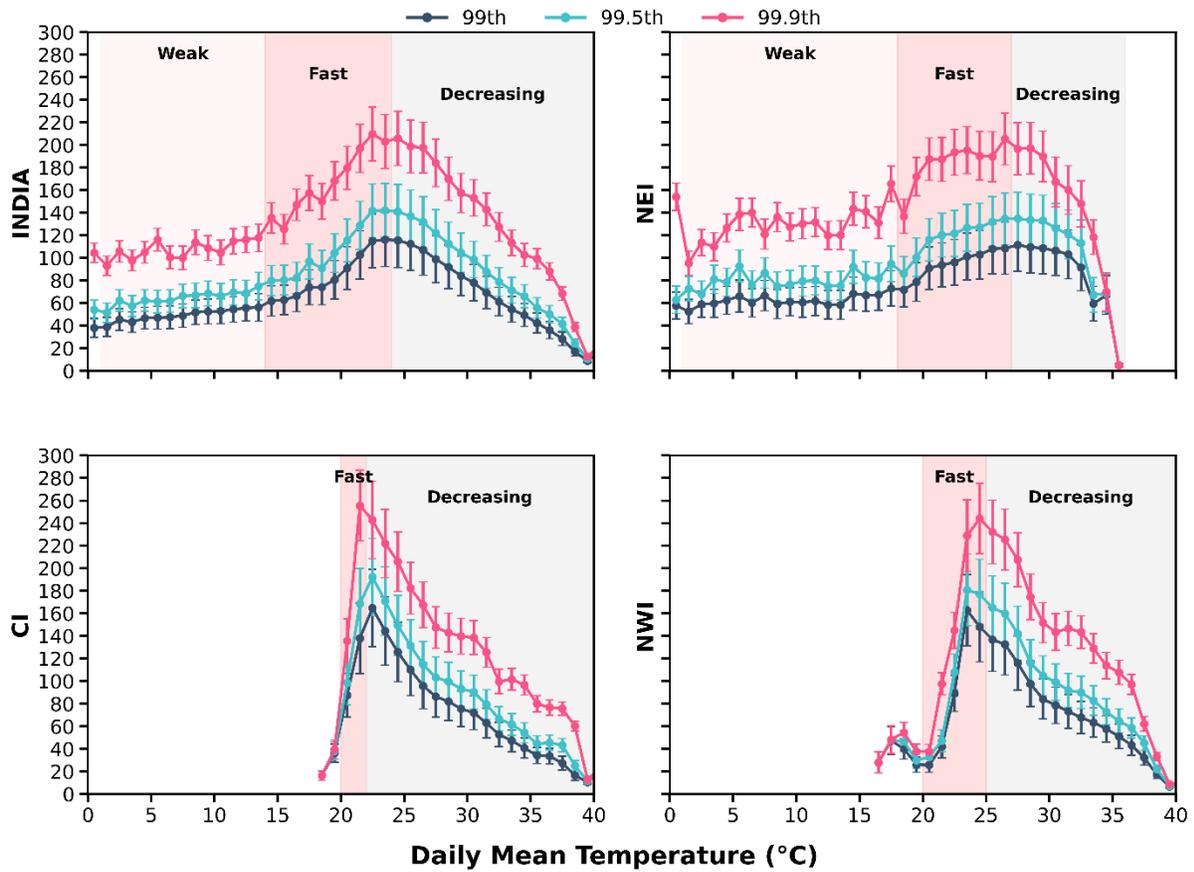

**Fig S2**
Temperature dependence of daily extreme precipitation intensity over India and its major sub-regions during the summer season using CMORPH precipitation and ERA5 2 m temperature data at 0.25° resolution. Shown are the 99th, 99.5th, and 99.9th percentile average precipitation intensities as a function of daily mean temperature (1 °C bins). The left column corresponds to summer extremes daily accumulated intensities for geographic domains: All-India, Northeast India (NEI), Central India (CI), and North-West India (NWI). Shaded backgrounds denote characteristic temperature regimes—Weak, Fast, and Decreasing—revealing nonlinear shifts in extreme rainfall behavior with temperature.

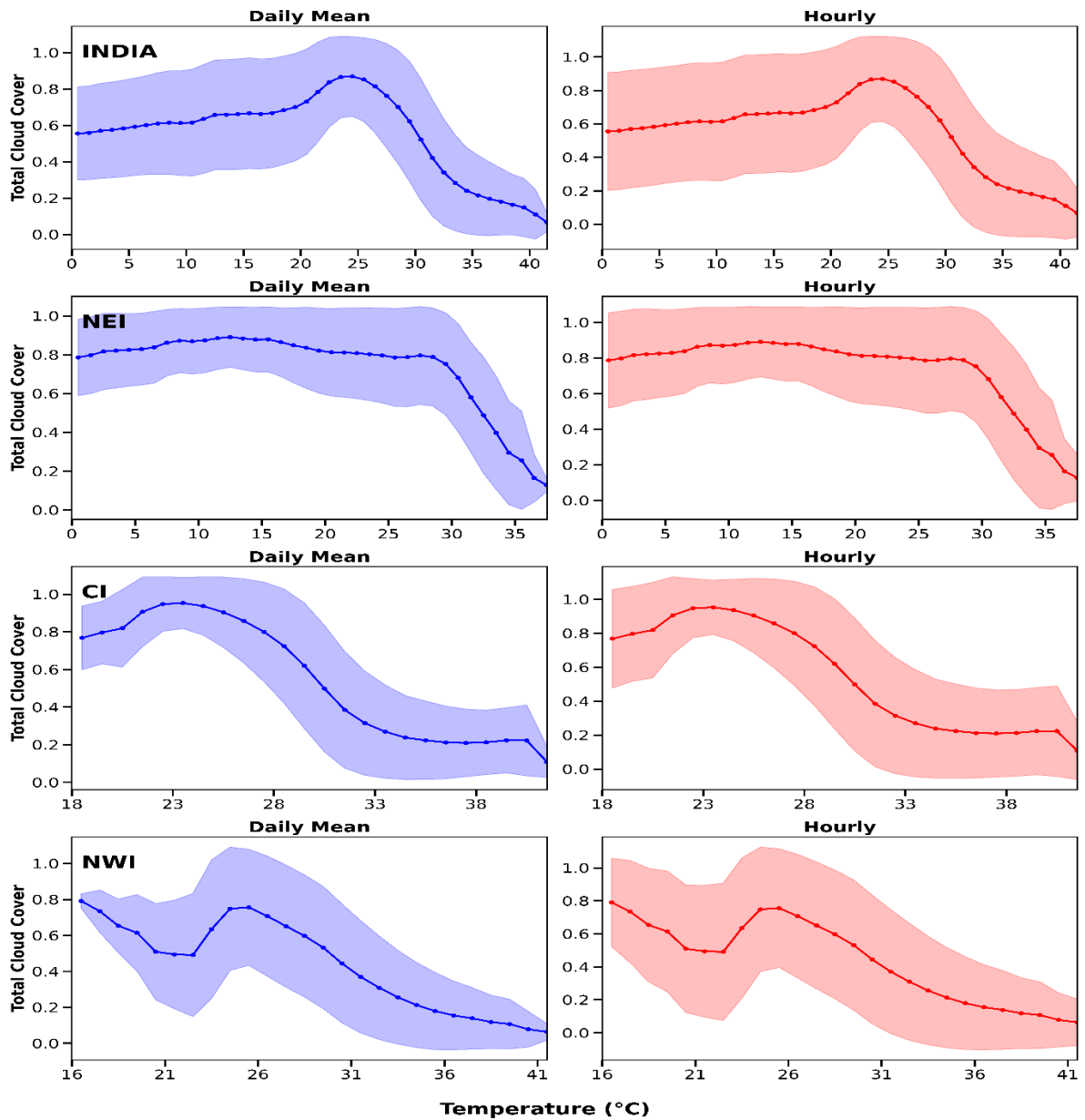

**Fig S3**
Temperature and cloud dependence over India and its major sub-regions during the summer season using ERA5 daily mean 2 m temperature and total cloud fraction at 0.25° resolution. Shown are the 99.5th percentile conditions as a function of daily mean temperature (1 °C bins) and total cloud fraction. The left column shows daily mean cloud fraction as a function of daily mean temperature at summer conditions for All-India, Northeast India (NEI), Central India (CI), and North-West India (NWI). The right column shows hourly total cloud fraction as a function of daily mean temperature, with shading indicating the standard deviation at daily and hourly scales.

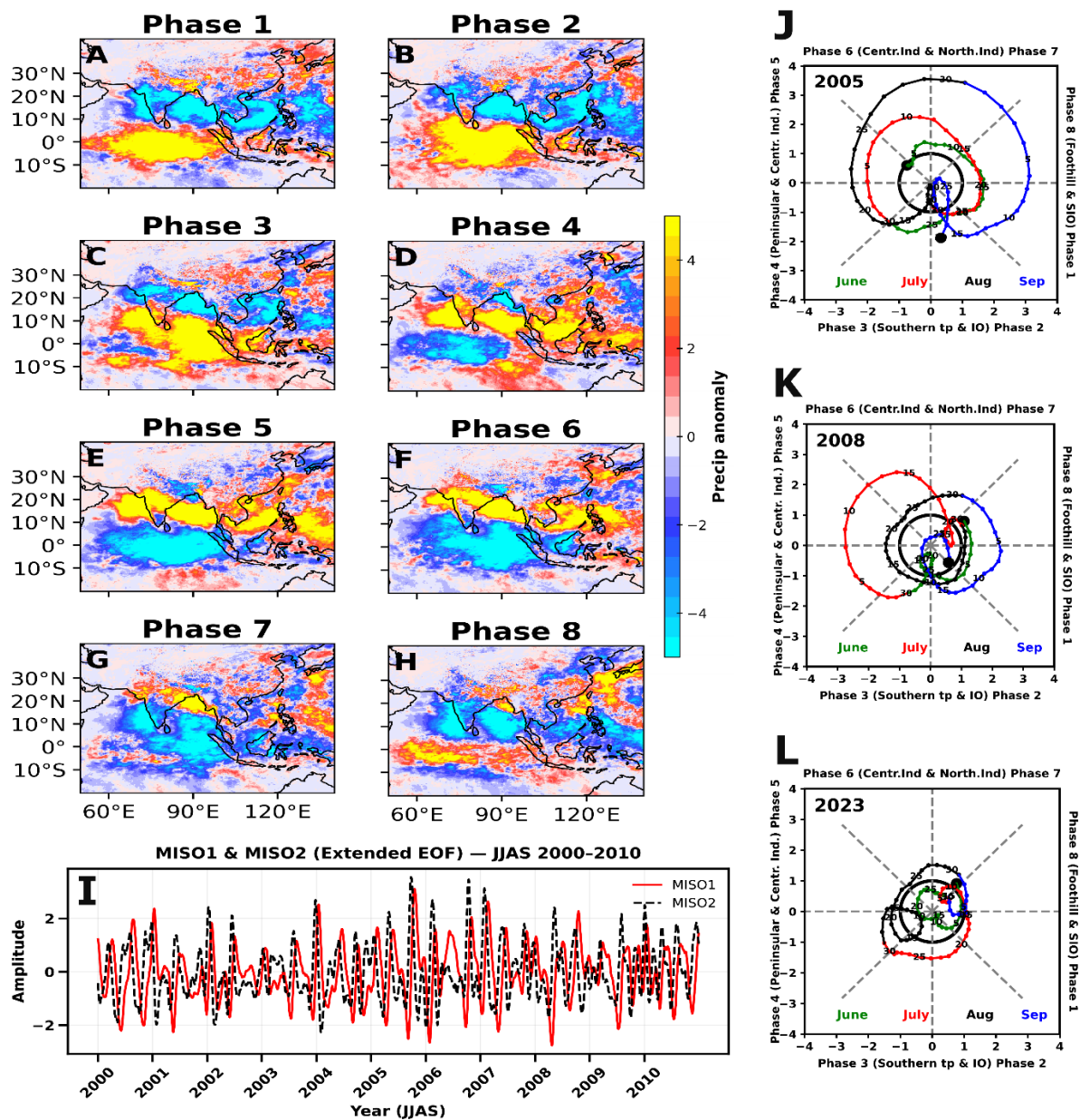

**Fig. S4.**

Panels A-H depict the different MISO phases over the ISMR region. Panel I. shows the first and second principal components (MISO1 and MISO2) of the extended EOF analysis derived from CMORPH datasets with a 15-day lag. Panels J-L present the MISO phase-space diagrams constructed using MISO1 and MISO2 as the x- and y-axes for the years 2005, 2008, and 2023, respectively.

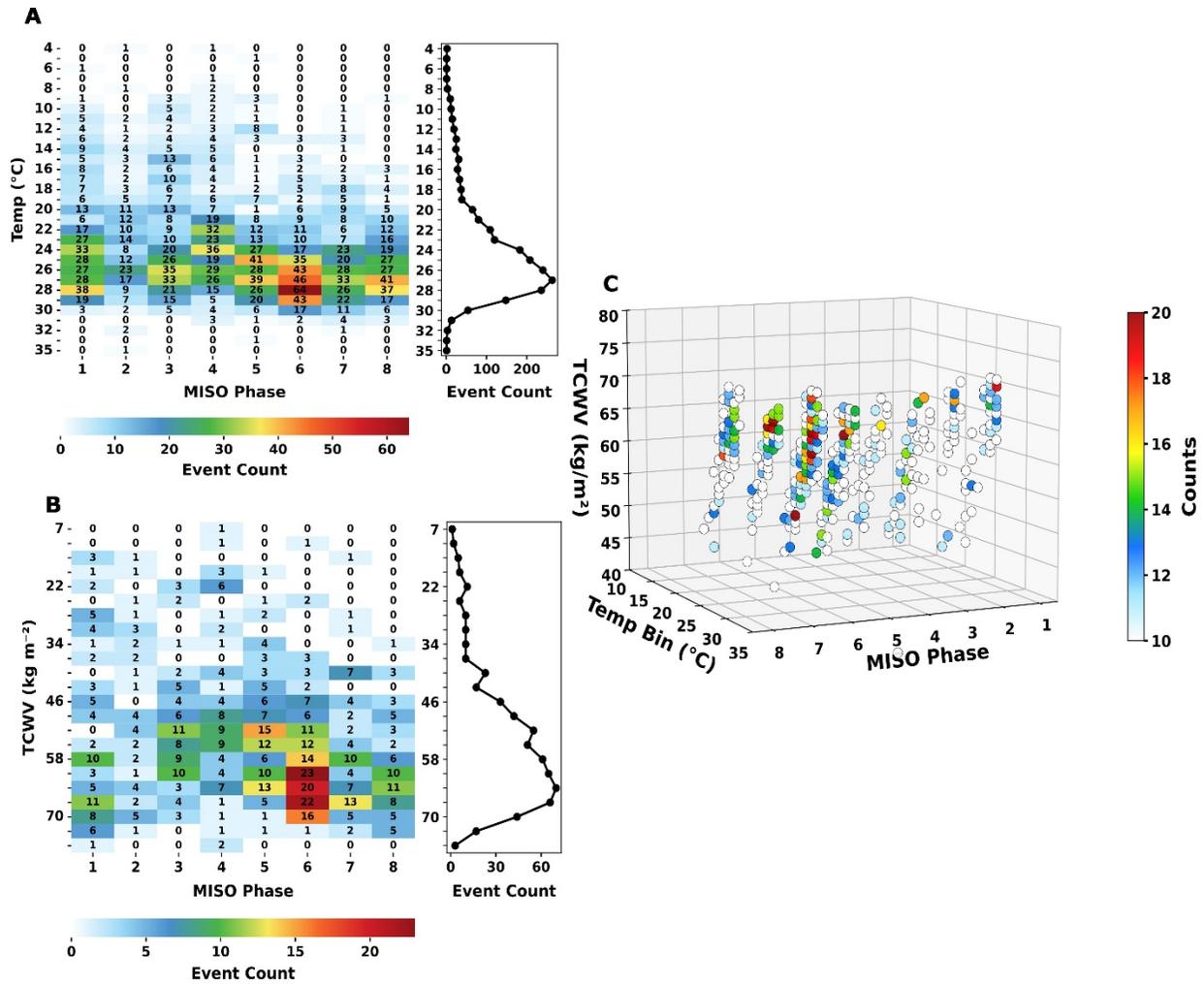

**Fig. S5.**
Bivariate and trivariate frequency distributions of 99.9th percentile extreme rainfall over NEI with respect to MISO phase and thermodynamic conditions. A. Bivariate distribution of extreme rainfall events as a function of MISO phase (x-axis) and daily mean temperature (y-axis); color shading indicates the count of events in each bin. B. Same as A., but with daily mean total column water vapor (TCWV) on the y-axis. C. Trivariate distribution showing the joint relationship between MISO phase, daily mean TCWV, and daily mean temperature; color intensity (or point density) represents the frequency of extreme rainfall occurrences within each 3D bin.

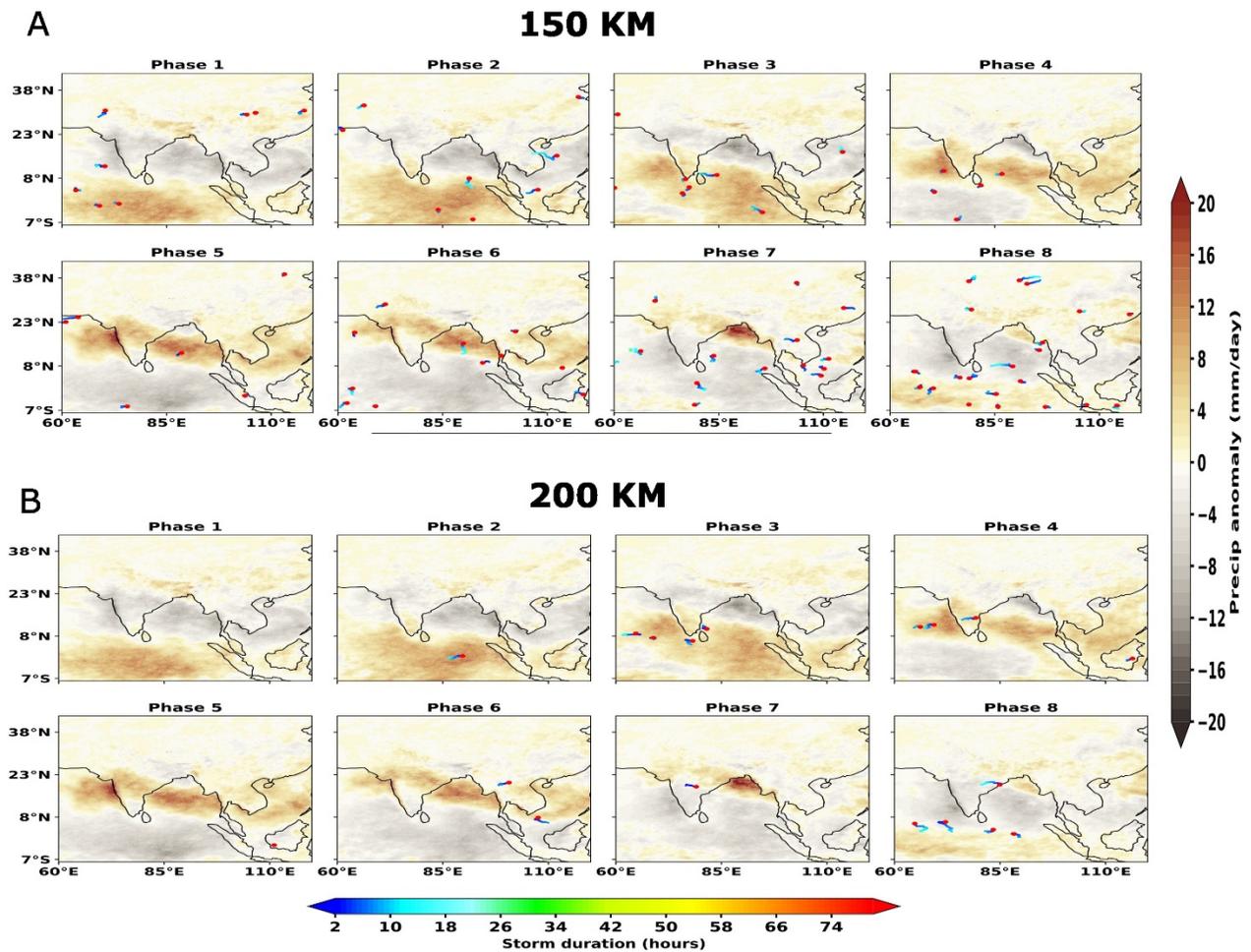

**Fig. S6.**

tracks of HERE storm cores generated in 2022, overlaid on MISO phases derived from the 1998–2023 climatological cycle. Panel A. represents 150-km storm cores, and panel B. represents 200-km storm cores.

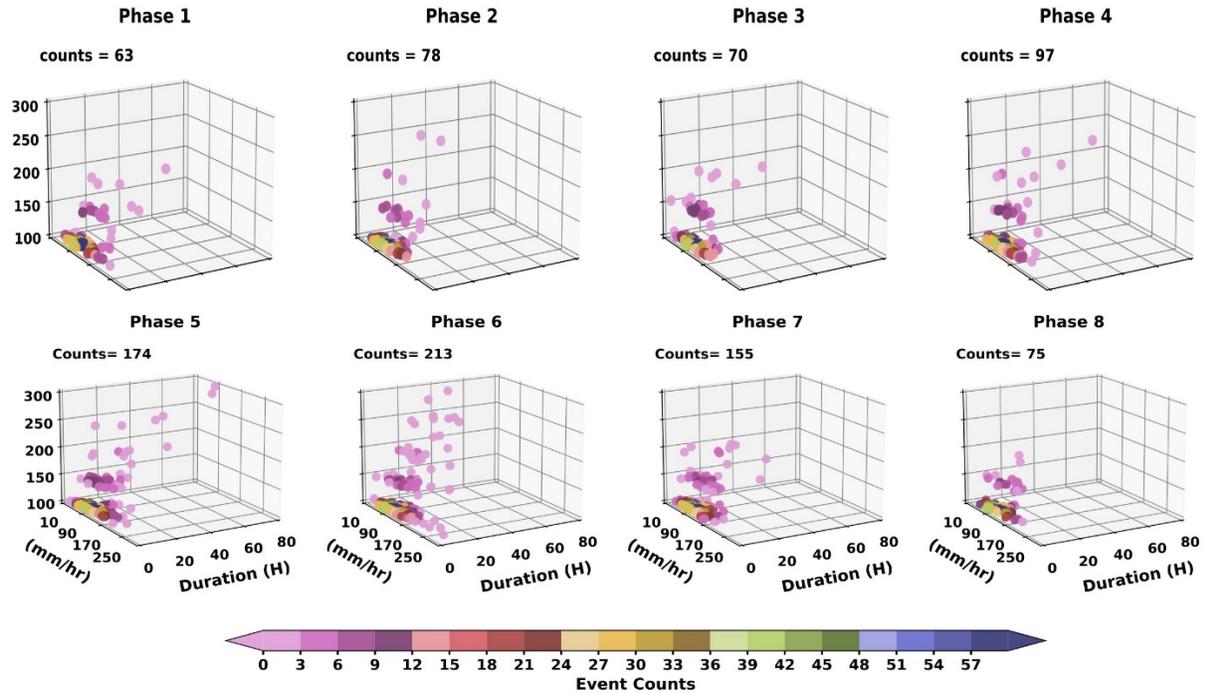

**Fig. S7.**
Frequency, duration, and intensity distributions of HERE storm cores classified by system size (100–300 km) across the eight MISO phases for the period 1998–2023. The figure illustrates storm occurrence, lifetime, and rainfall extremity vary systematically with storm scale and the evolving intraseasonal monsoon environment.

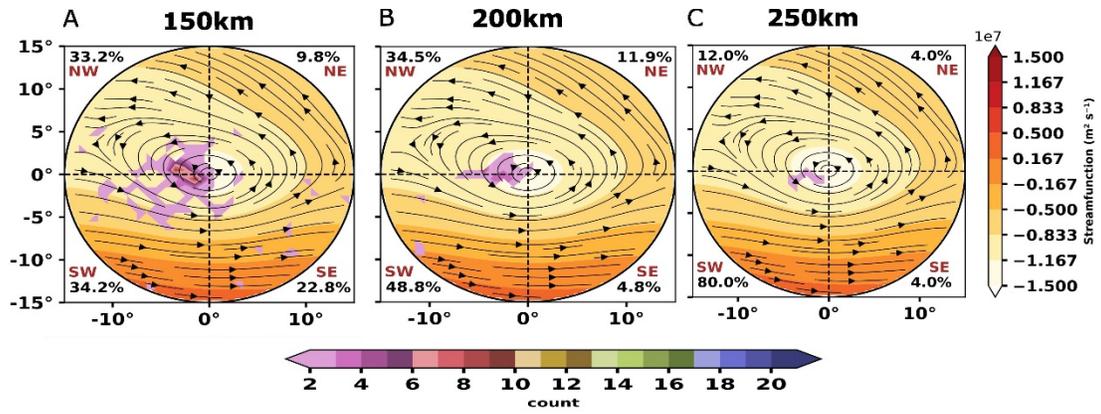

**Fig.S8.**
Composite structure of monsoon LPS divided in 4 quadrant from the center location overlaid by counts of HERE sizes from 150-250 km

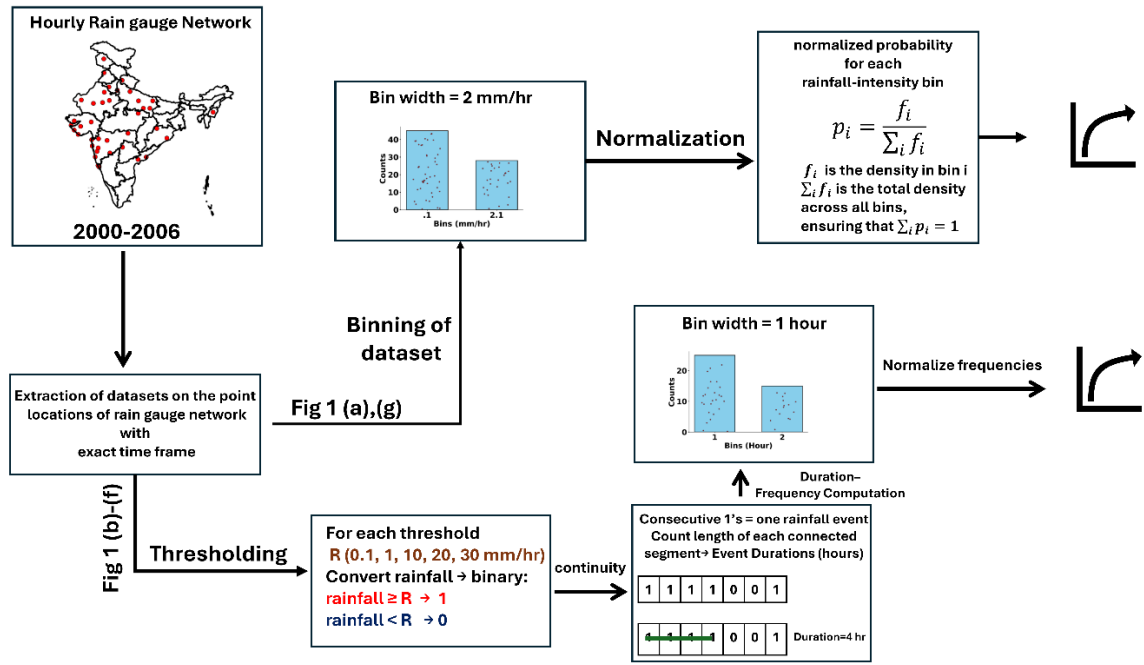

**Fig.S9.**
Schematic of the methodology used to analyze sub-daily rainfall characteristics at IMD rain gauge locations, showing the construction of hourly rainfall intensity distributions and rainfall spell durations from collocated hourly precipitation data.

| Region | Scale | 99.5 | | | 99.9 | | |
|---|---|---|---|---|---|---|---|
| | | Fast | Weak | Decreasing | Fast | Weak | Decreasing |
| INDIA | Daily | T:14–24<br>P:79.81–142.09<br>+7.80%/K | T:1–14<br>P:51.46–74.92<br>+3.51%/K | T:24–40<br>P:140.98–10.66<br>−5.78%/K | T:14–24<br>P:135.39–203.07<br>+5.00%/K | T:1–14<br>P:93.07–117.73<br>+2.04%/K | T:24–40<br>P:205.87–12.55<br>−5.87%/K |
| | Hourly | T:14–29<br>P:13.89–25.80<br>+5.72%/K | T:1–14<br>P:10.10–12.56<br>+1.88%/K | T:29–40<br>P:26.04–4.46<br>−7.54%/K | T:14–29<br>P:23.82–39.02<br>+4.26%/K | T:1–14<br>P:17.11–20.24<br>+1.41%/K | T:29–40<br>P:39.20–7.81<br>−7.28%/K |
| NEI | Daily | T:18–27<br>P:86.02–134.62<br>+6.28%/K | T:1–18<br>P:72.68–94.70<br>+1.78%/K | T:27–36<br>P:134.75–4.93<br>−10.70%/K | T:18–27<br>P:136.79–205.42<br>+5.57%/K | T:1–18<br>P:95.35–165.70<br>+4.34%/K | T:27–36<br>P:196.71–4.99<br>−10.83%/K |
| | Hourly | T:19–32<br>P:21.65–40.73<br>+6.78%/K | T:1–19<br>P:13.67–19.66<br>+2.44%/K | T:32–36<br>P:36.92–4.61<br>−21.88%/K | T:19–32<br>P:34.76–56.28<br>+4.76%/K | T:1–19<br>P:22.68–30.04<br>+1.80%/K | T:32–36<br>P:49.64–4.76<br>−22.60%/K |
| CI | Daily | T:20–22<br>P:98.40–168.57<br>+35.66%/K | – | T:22–40<br>P:192.25–11.48<br>−5.22%/K | T:20–22<br>P:135.64–255.53<br>+44.19%/K | – | T:22–40<br>P:242.79–12.91<br>−5.26%/K |
| | Hourly | T:19–26<br>P:9.11–28.50<br>+30.41%/K | – | T:26–40<br>P:29.03–5.42<br>−5.81%/K | T:19–26<br>P:10.18–40.30<br>+42.26%/K | – | T:26–40<br>P:40.55–9.09<br>−5.54%/K |
| NWI | Daily | T:20–25<br>P:32.80–176.90<br>+87.88%/K | – | T:25–40<br>P:165.06–7.41<br>−6.37%/K | T:20–25<br>P:37.45–244.13<br>+110.39%/K | – | T:25–40<br>P:232.21–8.72<br>−6.42%/K |
| | Hourly | T:20–27<br>P:9.41–33.06<br>+35.89%/K | – | T:27–40<br>P:34.60–5.43<br>−6.48%/K | T:20–27<br>P:11.54–46.85<br>+43.74%/K | – | T:27–40<br>P:48.70–5.98<br>−6.75%/K |

**Table S1.**

Thermal scaling of intensity changes percentage wise with temperature changes over 99.5 and 99.9 percentile